# Comparative study of magnetic and magnetotransport properties of $Sm_{0.55}Sr_{0.45}MnO_3$ thin films grown on different substrates

Manoj K. Srivastava[1,2], Sandeep Singh[1], P. K. Siwach[1], Amarjeet Kaur[2], V. P. S. Awana[1], K. K. Maurya[1], and H. K. Singh[1#]

[1]*National Physical Laboratory (Council of Scientific and Industrial Research), Dr. K. S. Krishnan Marg, New Delhi-110012, India*
[2]*Department of Physics and Astrophysics, University of Delhi, Delhi-110007, India*

**Abstract**

Highly oriented polycrystalline $Sm_{0.55}Sr_{0.45}MnO_3$ thin films (thickness ~100 nm) deposited on $LaAlO_3$ (LAO, (001)), $SrTiO_3$ (STO, (001)) and $(La_{0.18}Sr_{0.82})(Al_{0.59}Ta_{0.41})O_3$ (LSAT, (001)) single crystal substrates by ultrasonic nebulized spray pyrolysis have been studied. The out of plane lattice parameter (OPLP) of the film on LAO is slightly larger than that of the corresponding bulk. In contrast, the OPLP of the films on STO and LSAT are slightly smaller than the corresponding bulk value. This suggests that the film on LAO is under compressive strain while LSAT and STO are under tensile strain. The films on LAO and LSAT show simultaneous paramagnetic-ferromagnetic (PM–FM) and insulator-metal transition (IMT) temperature at $T_C/T_{IM}$ ~165 K and 130 K, respectively. The PM–FM and IM transition occur at $T_C$~120 K and $T_{IM}$~105 K, respectively in the film on STO substrate. At $T<T_C$, the zero field cooled–field cooled (ZFC–FC) magnetization of all the films shows strong bifurcation. This suggests the presence of a metamagnetic state akin to cluster glass formed due to coexisting FM and antiferromagnetic–charge order (AFM–CO) clusters. All the films show colossal magnetoresistance but its temperature and magnetic field dependence are drastically different. The films on LAO and STO show peak CMR around $T_C/T_{IM}$, while the film on LSAT shows MR>99 % over a very wide temperature range of ~40 K centred on $T_C/T_{IM}$. In the lower temperature region the magnetic field dependent isothermal resistivity also shows signature of metamagnetic transitions. The observed results have been explained in terms of the variation of the relative fractions of the coexisting FM and AFM–CO phases as a function of the substrate induced strain and oxygen vacancy induced quenched disorder.

[#] Author to whom correspondence should be addressed; Email: hks65@nplindia.org

**Introduction**

In doped rare earth manganites of the type $RE_{1-x}AE_xMnO_3$ (RE: rare earth cations; $La^{3+}$, $Nd^{3+}$, $Sm^{3+}$ etc., AE: alkaline earth cations; $Ca^{2+}$, $Sr^{2+}$ etc.) the lowering of the average RE/AE-site cationic radius ($\langle r_A \rangle$) decreases the $e_g$ electron bandwidth (W) that in turn results in increased carrier localization through the Jahn–Teller (JT) distortion of the $MnO_6$ octahedra. At reduced W the magnetic and magnetotransport properties show strong sensitivity to even weak external perturbations like small magnetic field, electric field, substrate induced strain, electromagnetic radiation, etc. and intrinsic disorders like oxygen and cationic vacancies, etc.[1-6] This is believed to be due to the enhanced competition between the ferromagnetic double exchange (FM–DE) that increases the kinetic energy of the itinerant $e_g$ electron and hence favours carrier delocalization and the JT distortion that favours antiferromagnetic superexchange (AFM–SE) and carrier localization.[1,2,7,8] Hence, at reduced W, the possibility of magneto-electric phase coexistence, especially in the vicinity of the half doping appears as a natural tendency. The most prominent example in this regards is $Sm_{1-x}Sr_xMnO_3$. This compound is unique due to its proximity to the charge order/orbital order (CO/OO) instability and shows the most abrupt insulator metal transition (IMT) and the most prominent magnetocaloric effect.[9-15] The ground states of $Sm_{1-x}Sr_xMnO_3$ are (a) ferromagnetic metallic (FMM) for $0.3<x\leq0.52$, and (b) antiferromagnetic insulating (AFMI) for $x>0.52$.[9-11] The charge ordering (CO) occurs in the range $0.4\leq x\leq0.6$ and the corresponding ordering temperature ($T_{CO}$) increases from ~140 to 205 K with $x$ increasing in the above range. Colossal magnetoresistance (CMR) is observed at all the compositions corresponding to the FMM ground state. Near half doping ($0.45\leq x\leq0.52$), very sharp (first order) transitions from paramagnetic insulating (PMI) to the FMM state are observed. Several studies on narrow band manganites have shown that the first-order nature of phase transition can be preserved even in presence of quenched disorder arising due to the size mismatch between RE and AE ions.[1] Like other low bandwidth manganites, $Sm_{1-x}Sr_xMnO_3$ has a natural tendency towards phase separation/phase coexistence (PS/PE) that causes evolution of a strong metamagnetic component around half doping ($x\sim0.50$). This metamagnetic makes the composition-temperature ($x$–T) phase diagram extremely fragile to external perturbations.

Despite detailed studies on the bulk polycrystalline and single crystalline $Sm_{1-x}Sr_xMnO_3$ forms, thin films have not been investigated in much detail. In this regard we would like to mention that as compared to the large and intermediate W manganites like $La_{1-x}Sr_xMnO_3$ and $La_{1-x}Ca_xMnO_3$ the growth of single crystalline $Sm_{1-x}Sr_xMnO_3$ thin films has been found to be rather difficult.[16-19] One of the factors that could supress the occurrence of PM-FM and IM transitions could be the extreme sensitivity of the magneto-electric phases (e.g., PMI, FMM and AFM–CO insulator (AFM–COI)) to the substrate induced strain in low W compounds. In small W compounds, (i) the reduced average RE-site cationic radius and hence smaller tolerance factor (t), (ii) the smaller Mn–O–Mn bond distance and the corresponding angle and (iii) enhanced size mismatch induced quenched disorder ($\sigma^2$) could lead to stronger sensitivity to the impact of substrate induced strain. As regards the impact of substrate it is generally accepted that the compressive (tensile) strain favours FMM (AFM–COI) and is inimical to the AFM–COI (FMM) phases.[6,20] However, in narrow W manganites like $Sm_{1-x}Sr_xMnO_3$ substrate induced strain may not be the only factor determining the magneto-electric phase profile. It is worth mentioning that there are some reports showing the anomalous behaviour wherein the compressively strained thin films on LAO substrates do not show any IMT at lower film thickness (e.g., 25 nm and 50 nm), which, however, is seen in 120 nm film.[21] In contrast the tensile strained film on STO shows IMT even at film thickness of 50 nm.[21] The impact of substrate induced strain on magnetic phase coexistence and consequent magnetotransport properties of small W material like $Sm_{1-x}Sr_xMnO_3$ appears to be more dramatic.[19,22] In fact recently it has been shown that even a small strain can cause appreciable modifications in the magnetoelectric phase landscape of a low W manganite like $Sm_{1-x}Sr_xMnO_3$.[23]

Recently we have studied $Sm_{0.55}Sr_{0.45}MnO_3$ (SSMO) thin films, wherein it was demonstrated that the substrate induced strain and oxygen vacancy ordering/disordering have significant impact on the magnetotransport properties.[23] In the continuation of above mentioned work, here we report the detailed study on SSMO thin films grown over LAO, STO, and LSAT substrates, which provide compressive strain, tensile strain and least strain, respectively. Our results clearly demonstrate that despite the polycrystalline nature of these films the impact of substrate is indeed dramatic and unambiguously manifested in the magnetic and magnetotransport properties.

**Experimental Details**

Polycrystalline thin films of $Sm_{0.55}Sr_{0.45}MnO_3$ (thickness ~100 nm) on single crystals LAO (001), STO (001) and LSAT (001) substrates were synthesized by using ultrasonic nebulized spray pyrolysis.[23] Stoichiometric amounts of high purity Sm, Sr, and Mn nitrates (Sm/Sr/Mn=0.55/0.45/1) were dissolved in deionized water and the solution was homogenized. Film deposition was done at substrate temperature; $T_S$~200 °C and the films were annealed in air at temperature $T_A \sim 1000\,°C$ for 12 hrs, followed by slow cooling with cooling rate 4 °C/min. Here we would like to point out that high temperature annealing does not lead to any observable interdiffusion at the film substrate interface.[24] The structural and surface characterizations were performed by X-ray diffraction (XRD, PANalytical PRO X'PERT MRD, Cu-$K_{\alpha 1}$ radiation $\lambda=1.5406$ Å) and atomic force microscopy (AFM), respectively. The cationic composition was studied by energy dispersive spectroscopy (EDS) attached to scanning electron microscope. The temperature and magnetic field dependent magnetization was measured by a commercial (Quantum Design) PPMS at H=500 Oe magnetic field applied parallel to the film surface. The electrical resistivity was measured by the standard four probe technique in the magnetic field range $0 \leq H \leq 50$ kOe.

**Results and Discussion**

The XRD data (Fig. 1) shows the occurrence of the (00$\ell$) reflections alongside the corresponding substrate peaks (marked by S in Fig. 1) and absence of any other diffraction maxima corresponding to the film material. This shows strong texturing and orientation along the out of plane direction. The crystal structure of SSMO ($x$=0.45), as reported by Tomioka et al.,[9] is orthorhombic (Pbnm) and the c-parameter corresponding to the cubic unit cell is c ≈ 3.83 Å. The out of plane lattice parameter (OPLP) of the film on LAO substrate is $c_{LAO}$=3.855 Å, which is found to be larger than the corresponding bulk value. The OPLP of films on STO and LSAT are $c_{STO}$=3.822 Å and $c_{LSAT}$=3.826 Å, respectively. These estimations suggest that the films grown on LAO are compressively strained ($a_{LAO}$=3.79 Å) and slightly smaller lattice constants of SSMO on STO ($a_{STO}$=3.905 Å) and LSAT ($a_{LSAT}$=3.868 Å) substrates could be attributed to the small tensile strain. As mentioned earlier the tensile strain is believed to favour the AFM-COI phase, while the compressive strain enhances the FMM fraction. At the lattice level the compressive strain results in an elongation of the $MnO_6$ octahedra in the OP direction with a concomitant compression in the basal plane that causes a reduction in the degree of JT distortion, and hence weakens the spin-lattice coupling. On the other hand the tensile strain elongates the $MnO_6$ octahedra in the basal plane with a concomitant compression along the OP direction.[6,20] Here, we must point out that the impact of strains in polycrystalline films is expected to be of localized character (due to the presence of discontinuity at the grain boundaries, where the strain could be relaxed easily) and hence may not be unambiguously visible (as in case of epitaxial/single-crystalline thin films) in gross structural characteristics like XRD patterns. As revealed by SEM (not shown here), the surface of these films generally consists of a mixture of large continuous layers, which are intermittently covered by small granules. This could be suggestive of local epitaxial like grown regions having strong texturing. Surface topography of all the films was probed by AFM. In case of the film on LAO the surface appeared to consist of large and big granules, while in case of the films on LSAT and STO substrates the granule size are more uniform. As compared to single crystalline $Sm_{0.53}Sr_{0.47}MnO_3$ thin films prepared by DC magnetron sputtering,[19,22] the surface roughness of these polycrystalline films is relatively higher. The representative surface topographs of films on LAO and STO are presented in Fig. 2.

The temperature dependent resistivity (ρ–T) of these films, measured at H=0 kOe and H=50 kOe, is plotted in Fig. 3. The zero filed insulator-metal transition (IMT) temperature ($T_{IM}$) of the SSMO films on LAO, LSAT and STO substrates are found to be ≈164 K, 130 K and 105 K, respectively. From the resistivity profile of the films it is clear that the IMT of the film on LAO is broadest and that of the film on LSAT substrate is the sharpest, where the resistivity decreases sharply by nearly three orders of magnitude. The sharpness of the IMT is also demonstrated by the temperature coefficient of resistivity (TCR) [defined as $TCR\ (\%) = \frac{d}{dT}(ln\rho)x100$], which is an important property from application. The peak TCR value of films on LAO, LSAT and STO is found to be ≈7 %, 31 % and 9 % respectively. Here we must point out that the film on LAO shows large enhancement in the transition temperature as compared to the polycrystalline/single crystalline bulk ($T_{IM}$ ~130 K) and thin films of similar composition.[9-17] However, the abrupt (first order) transition seen in such bulk poly- and single crystalline samples[1,9,10] has been transformed into a continuous second order transition in this film. Furthermore, the hysteretic behaviour of the ρ–T measured in heating-cooling cycles is also absent in this film.[23] The most probable reason for blocking of the first order phase transition in the film on LAO appears to be the presence of quenched disorder.[2,3] The films on LSAT and STO show irreversibility in ρ–T data (results not shown) that is more pronounced in the later. The difference in the $T_{IM}$ measured during the two cycles is up to ~10 K. The application of the magnetic field leads to decrease in the resistivity, enhancement in $T_{IM}$ and broadening of the transition. At H=50 kOe the IMT of films on LAO, LSAT and STO is enhanced to ≈200 K, 184 K and 128 K, respectively. The corresponding magnetic field induced enhancement in the IMT values ($\Delta T_{IM}$) are ≈36 K, 54 K and 23 K for films on LAO, LSAT and STO, respectively. This shows that magnetic field induced enhancements (10.8K/10 kOe) in the IMT is the largest in the film on LSAT.

The temperature dependent zero field cooled (ZFC) and field cooled (FC) magnetization data (M–T) measured at H=500 Oe magnetic field is shown in Fig. 4. All the films show well defined PM-FM transition and the Curie temperature is found to be $T_C \approx 165$ K, 130K and 120 K, respectively for LAO, LSAT and STO films. The magnetization of the LAO film starts rising at T≈190 K and then shows FM transition at $T_C \approx 165$ K, which like the IMT, is uncharacteristically broad for a low W compound like SSMO. At $T<T_C$, the ZFC and FC branches are observed to diverge appreciably. In the low temperature regime the ZFC magnetization ($M_{ZFC}(T)$) shows a cusp like feature at $T_P \approx$ 40 K (LAO), 45 K (LSAT) and 40 K (STO), and then drops sharply below this point. In the FC magnetization ($M_{FC}(T)$) curves, $T_P$ is shifted to lower temperature and the sharpness of the magnetization drop at $T<T_P$ is reduced considerably in all the films. The low temperature drop in magnetization, coupled with the ZFC–FC divergence is a signature of a metamagnetic state, most likely a cluster glass (CG).[23,25] The ZFC–FC divergence and sharpest magnetization drop observed in the LSAT films show that they possess the highest CG fraction. The occurrence of the metamagnetic states like CG could be explained in terms of the coexistence of the short range AFM–COI and A–AFM correlations in the vicinity of $T_C/T_{IM}$ in $Sm_{1-x}Sr_xMnO_3$ (x~0.5), which is well established in the bulk form of the compound.[1,9,10] However, in thin films the structural-microstructural modifications induced by substrate induced strain could lead to appreciable change in the magnetic landscape even at $T<T_C$. Hence, in the present case (i) long range FMM, (ii) short range AFM–COI and (iii) short range A–AFM phases are expected to coexist also at $T<T_C$. Such phase-coexistence could cause frustration in the FM phase and hence result in the formation of the CG at $T<T_C$. The variation in relative fraction of FMM and AFM–COI phases is also expressed in magnitude of the magnetization. In this regard it is clear from the magnetization data presented here that the film on LAO shows the highest magnetic moment, while lower magnetic moment has been found in the case of LSAT and STO films. This observed variation in magnetic moment could be attributed to different type of strains provided by the substrates. Since the compressive strain favours the FMM, the film on LAO possess the highest value of magnetic moment, while the smallest value of the magnetic moment in the film on STO is a consequence of the tensile strain, which favours AFM–COI. The representative M–H plot of all the films is shown in the inset of Fig. 4. The saturation moment ($M_S$) extracted from the M–H loop is ≈585 emu/cm$^3$, 525 emu/cm$^3$ and 362 emu/cm$^3$ for LAO, LSAT and STO, respectively and the corresponding magnetic field ($H_S$) is found to be ~4 kOe, ~7 kOe and ~8 kOe, respectively. The asymmetric coercivity, suggests the presence of exchange bias (EB) effect due to coexisting AFM and FM clusters. The coercivity asymmetry is found to be the smallest in the film on LAO and largest in the film on STO. Thus our results show that compressive strain in the film on LAO results in (i) enhanced $T_C/T_{IM}$, (ii) broadening of the PM–FM and IM transition, (iii) high saturation moment and lower saturation field, and (iv) lower coercivity. The tensile strain although very small, has the impact which is just opposite of the above.

The large enhancement in $T_C/T_{IM}$ of the films on LAO and the concomitant blocking of the abrupt resistive transition could be understood in terms of (i) the localized compressive strain and (ii) oxygen vacancy induced suppression of the AFM–COI state, which coexists with the PM ($T>T_C$) and FMM ($T<T_C$) phase. Recently, it has been shown that even a very small substrate induced strain could modify the magnetotransport properties in $Sm_{1-x}Sr_xMnO_3$ (x~0.5) in the FMM regime ($T<T_C$).[19,22] In case of textured polycrystalline thin films the nature of the strain state is expected to be drastically different from the single crystalline and epitaxial thin films. In the textured/oriented polycrystalline thin film the local epitaxial regions are interrupted by the presence of GBs and around these regions the strain, irrespective of its nature (whether it is compressive or tensile) is expected to get relaxed. Such strain discontinuity at the GBs would make the strain weak and spatially non uniform. These GBs works as inhomogeneity which could cause quenched disorder (QD).[1-3,23] Further, in manganites, it has been demonstrated that the oxygen vacancies can destabilize the AFM–COI phase both in single crystalline as well as polycrystalline materials quite efficiently.[23,26,27] Since the oxygen stoichiometry is related to the effective hole concentration and even a mild spatial inhomogeneity in the oxygen vacancies could result in spatially varying carrier density that may also act as QD. Thus the origin of the quenched disorder could be traced to (i) the compressive strain provided by the substrate, and (ii) the ordering of oxygen vacancies created by high temperature annealing. Such vacancies are expected to be more at and in the vicinity of the film surface and the film-substrate interface. Here we must emphasise that ordering and disordering of these oxygen vacancies could have the decisive role.[23] As we have earlier shown that the abrupt IMT, which is akin to a first order phase transition is recovered either by rapidly cooling these films after annealing in air, or annealing this film in flowing oxygen and then cooling it slowly.[23] Thus in textured/oriented polycrystalline thin films two types of quenched disorders could possibly arise, the first being due to the strain inhomogeneity and the second due to the oxygen vacancy induced carrier density inhomogeneity. Further, since highly oriented films the grain boundary contribution to the electrical transport properties is considerably reduced therefore the contribution of the QD is expected to be decisive. Thus in case of the film on LAO the QD could transform the long range AFM–COI into short range and enhance the FMM fraction. This explains the observed rise in the magnetization at T>$T_C$ as well as the blocking of the abrupt resistive transition. However, the films on STO show stronger decrease in $T_C/T_{IM}$, while LSAT films show same $T_C/T_{IM}$ as reported for polycrystalline/single crystalline bulk ($T_C/T_{IM}$ ~130 K) and thin films of similar composition. The huge decrease in $T_C/T_{IM}$ of STO films could be attributed to the substrate induced tensile strain which strengthens the JT distortion of the $MnO_6$ octahedra and hence favours the AFM–COI phase. Thus we can conclude that spatially inhomogeneous strain as well as other factors such as the oxygen vacancy, etc. also plays a crucial role in determining the magnetotransport properties in low W manganites.

The temperature dependence of MR measured at H=50 kOe of all the films is plotted in Fig. 5. In case of the film on LAO the MR rises rapidly on lowering the temperature and has a peak value of 87 % at T≈140 K. On further lowering the temperature the MR decreases and saturates to ≈40% at 5 K. The fact that the peak in the MR–T curve of the film on LAO occurs much below the $T_C/T_{IM}$ could also be a possible consequence of the coupled effect of quenched disorder and compressive strain. In the film on STO the MR rises very slowly till T≈ 160 K and then undergoes a sharp increase, reaching the peak value ≈91 % at T≈100 K. In the lower temperature region the MR of this film decays to ≈55 % at 5 K. As the temperature is lowered the temperature dependence of MR in the film on LSAT is similar to that in the film on LAO till T≈190 K. Below this temperature the MR rises sharply and approaches ≈99 % at T≈144 K. Interestingly the MR of this particular film remains in excess of 99% in the temperature range 144–110 K, hence causing a plateau like feature in the MR–T curve (Fig. 5.) Thus it is clear that the MR in the film on LSAT substrate approaches 99 % about 15 K above the $T_C/T_{IM}$ and remains nearly constant down to 110 K. The occurrence of CMR over such large temperature range suggests towards appreciable presence of AFM–COI cluster in this temperature range. These AFM–COI clusters are transformed into FMM ones by the applied magnetic field. At this point we would like to mention that this film also shows the largest shift in the $T_{IM}$ due to the applied magnetic field, which as mentioned earlier is $\Delta T_{IM}$ ≈54 K. The observed pattern in the variation of MR–T data could be related to the different ratios of the two competing magnetoelectric phases, viz. FMM and AFM–COI on different substrates.

Isothermal magnetic field dependent resistivity (ρ–H) measured at several temperatures shows many interesting features. In the lower temperature regime, e.g., T=5 K, the ρ–H data of all the films shows signature of a soft metamagnetic component. The normalized isothermal resistivity [ρ (H)/ρ (50 kOe)] is plotted as a function of the applied magnetic field in Fig. 6. As seen in the plot, in the initial magnetic field cycle the resistivity of all the films first shows slow decrease as the H is increased and then drops sharply beyond a critical magnetic field value $H^*$. This kind of feature is generally attributed to the collapse of the AFM–COI state, that is the magnetic field induced AFM-COI to FMM transformation.[28] The observed values of $H^*$ is ≈17.5 kOe, 22.5 kOe and 25 kOe in the film on LAO, LSAT and STO, respectively. This clearly shows that the AFM–COI state is the strongest in the film on STO substrate. In the subsequent cycles, although the initial value of the resistivity is not re-attained but in all the films the ρ–H curves show strong hysteresis. Except for the slope the ρ–H loop of the films on LAO and STO is nearly similar. In contrast, at certain values of H, the ρ–H loop of the film on LSAT shows sharp jumps that occur at different field values during the field increasing and decreasing cycles. The origin of such features is believed to be the occurrence of a metamagnetic component due to competing FMM and AFM–COI phases.[17,28] The sharp magnetic field drop in the resistivity is observed only in the lower temperature regions and is absent in the ρ–H loops measured at T≥50 K. Hence this could be correlated to the occurrence of a metamagnetic component as also evidenced by the strong bifurcations in the ZFC–FC magnetization curves of these films (Fig. 4). As explained earlier such bifurcation of the ZFC–FC curves is regarded as generic feature of the CG like metamagnetic state, which in the present case is caused by coexisting FMM and AFM–COI phases. As demonstrated by the magnetization and electrical transport data, the film on the LAO substrate has the lowest fraction of the AFM–COI phase. This explains the smallest value of the $H^*$ (≈17.5 kOe) in the film on LAO. On the contrary, the film on STO has the highest AFM–COI component and hence the value of $H^*$ is the largest in this. The sharp drop and the fact that the virgin resistivity is not achieved in the subsequent cycles suggests that the AFM ordered COI clusters are melted by the applied field and major fraction of these get transformed permanently into FMM ones, that is the AFM–COI to FMM transformation is not fully reversible.

The MR calculated from the isothermal resistance measurements is plotted in Fig. 7 (a–f). At T=5 K, the film on LAO and STO have almost identical behaviour, wherein they show similar hysteresis and rather small MR~13% at 50 kOe. In contrast the MR of film on LSAT shows jump, in addition to hysteresis, that occurs at different magnetic fields during the field increasing and decreasing cycles. Beyond these jumps, the slope of the MR–H curves is changed, albeit no saturation of the MR is observed up to 50 kOe. At 50 K, (Fig. 7b) the film on LAO shows the lowest MR≈39 % at H=50 kOe and it also has the narrowest hysteresis. The film on STO shows MR≈51 % at H=50 kOe and a strong hysteresis is seen in the MR-H curve. In the film on LSAT the MR increases sharply as H is increased up to ~20 kOe and beyond that the slope of the MR–H curve is appreciably lowered but no saturation like behaviour is seen. At T=100 K, the film on LAO shows MR≈72 % (H=50 kOe) and the hysteretic behaviour of the MR–H curve has almost vanished. The films on STO and LSAT, in contrast still show strong hysteretic field dependence and much higher MR. The MR in both these films rises sharply up to H≈20 kOe. However, MR in none of these films shows a saturation tendency up to H=50 kOe. At T= 125 K, the hysteresis in the MR–H curve of the film on LAO vanishes and MR shows a sharp rise at H ≤ 20 kOe. In this film the saturation tendency is still not seen. The MR of the film on LSAT still shows a strong hysteresis and sharply reaches ~99 % at H≈20 kOe and saturates at slightly higher fields. Since the measurement temperature (125 K) is higher than the $T_C/T_{IM}$ of the film on STO (PM phase), the MR in this film remain very small till H≈15 kOe and then rises sharply with weak saturation like tendencies appearing around H=50 kOe. At T=150 K, in all the films the MR–H hysteresis vanishes. Since the film on LAO is still in the FM state, its MR is still about ≈81 % at 50 kOe. The film on STO does not have any significant MR till about H=25 kOe but at further higher fields, MR approaches ~18 %. In case of the film on LSAT, the MR rises very sharply beyond H≈10 kOe and appear to saturate at ~96 % at H=50 kOe. The occurrence of such large MR at T > $T_C$ and the nonlinear nature of the MR–H curve clearly suggests that AFM–COI cluster could be present in the PM regime also and as the magnetic field is increased they get transformed in to the FMM. At T=200 K, all the films show typical behaviour of MR in the paramagnetic regime above IMT having linear increment of MR which decreases to ≈47 %, 33 % and 4 %, respectively for LAO, LSAT and STO films. The MR–H data presented above clearly suggest that the temperature dependent hysteresis could be consequence of the varying fractions of the FMM and AFM–COI phase at T < $T_C$ and AFM–COI and PMI phases at T > $T_C$. The occurrence of very large hysteretic MR in the film on LSAT even at moderate magnetic fields at T < $T_C$ is clear signature of the presence of AFM–COI clusters in the FMM regime, while the presence of non-hysteretic but nonlinear MR at T > $T_C$ shows the presence of AFM–COI clusters in the PMI regime. One more aspect for the occurrence of hysteresis in MR–H curve could be the nature of magnetic spin alignment in both directions of magnetic fields. In increasing magnetic fields the spins are easily aligned in field direction but when reverse magnetic field is applied it requires more energy to rotate magnetic spins in the field direction to achieve previous magnitude of magnetoresistance indicating a strong coupling between spin and magnetic easy axis.

**Conclusion**

In summary, we have synthesized oriented high quality polycrystalline SSMO thin films deposited on LAO, LSAT and STO single crystal substrates and investigated the impact of substrate on magnetic and magnetotransport properties. Our results clearly show that even a subtle change in the nature and magnitude of the strain results in appreciable modifications in the magnetic and magnetotransport properties. The large enhancement in the $T_C/T_{IM}$ of the film on LAO with a simultaneous blocking of the abrupt resistive transition has been explained in terms of quenched disorder whose origin has been traced to the inhomogeneous compressive strain and the surface/interfacial oxygen vacancies. In contrast, the decrease in the $T_C/T_{IM}$ of the film on STO has been caused by the enhanced AFM–COI fraction due to the tensile strain. The film on LSAT, which is least strained shows the sharpest PM–FM and an abrupt insulator-metal transitions, the strongest metamagnetic component and CMR in excess of 99% over a broad temperature around $T_C/T_{IM}$. This shows that the coexistence of the FMM and AFM–COI phases is more delicately balanced in film on LSAT. The competing FMM and AFM–COI phases cause a metamagnetic state akin to the cluster glass. The substrate and temperature dependent variation in the fraction of the FMM and AFM–COI phases has a strong bearing on the magnetotransport properties on these films.

**Acknowledgement**

MKS is thankful to CSIR, New Delhi for research fellowship. Authors at CSIR–NPL acknowledge the continued institutional support through the in-house project (OLP#120632) and thank Prof. R. C. Budhani.

**Figure Captions**

Fig. 1: X-ray diffraction patterns (2θ/ω scan) of the $Sm_{0.55}Sr_{0.45}MnO_3$ films on LAO, LSAT and STO substrates. Substrate peaks are marked as S and $Sm_{0.55}Sr_{0.45}MnO_3$ reflections are indexed.

Fig. 2: The representative AFM surface topographs of the $Sm_{0.55}Sr_{0.45}MnO_3$ films on (A) LAO and (B) STO substrates.

Fig. 3: Temperature dependence of resistivity of the $Sm_{0.55}Sr_{0.45}MnO_3$ films measured in the range 4.2–300 K at H=0 and H= 50 kOe.

Fig. 4: Temperature dependent ZFC & FC magnetization (H=500 Oe) of all the $Sm_{0.55}Sr_{0.45}MnO_3$ films. Inset shows the M–H loops of all the films measured at 5 K.

Fig. 5: Temperature dependence of magnetoresistance of the $Sm_{0.55}Sr_{0.45}MnO_3$ films measured at H=50 kOe.

Fig. 6: Variation of resistivity as a function of the magnetic field measured at 5 K. The arrows in the upper part of the figure indicate the critical magnetic field ($H^*$) beyond which the resistivity drops sharply. The arrows along with the LSAT data mark the direction of the magnetic field cycling and the same is valid for all the samples.

Fig. 7(a–f): Isothermal magnetic field dependent magnetoresistance (MR–H) of LAO, LSAT and STO films measured at (a) 5 K, (b) 50 K, (c) 100 K, (d) 125 K, (e) 150 K and (f) 200 K.

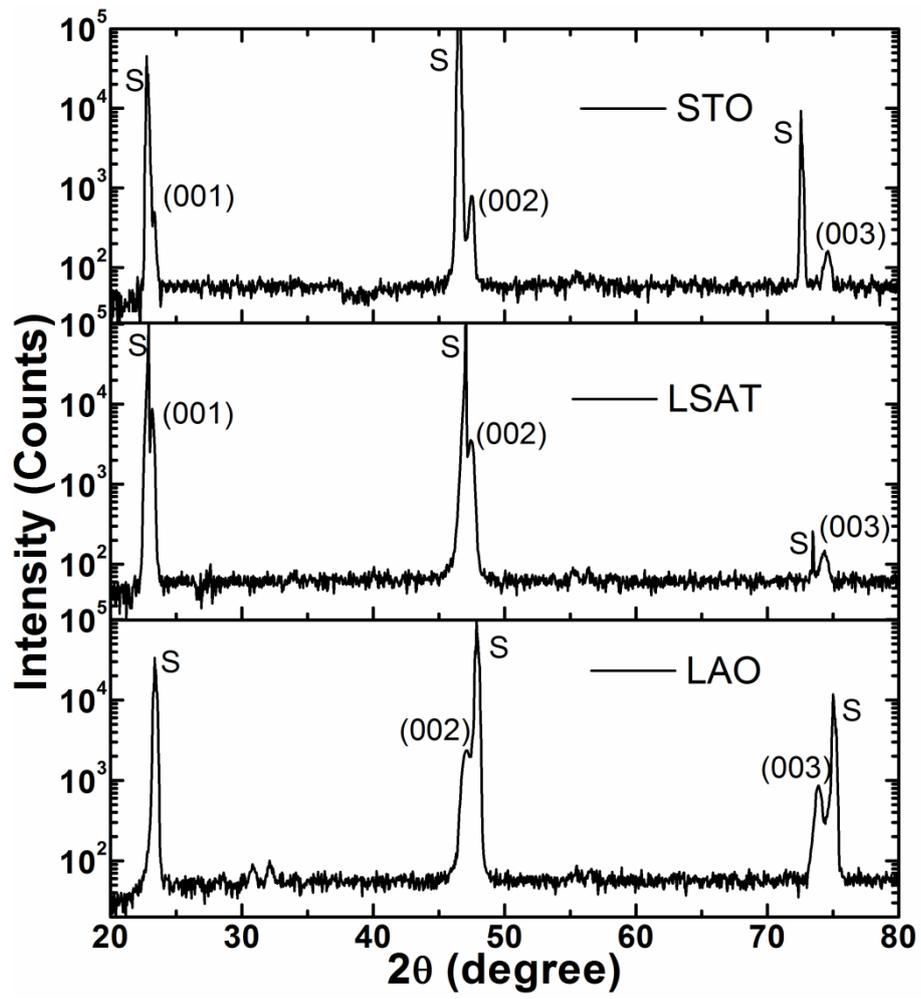

Fig. 1

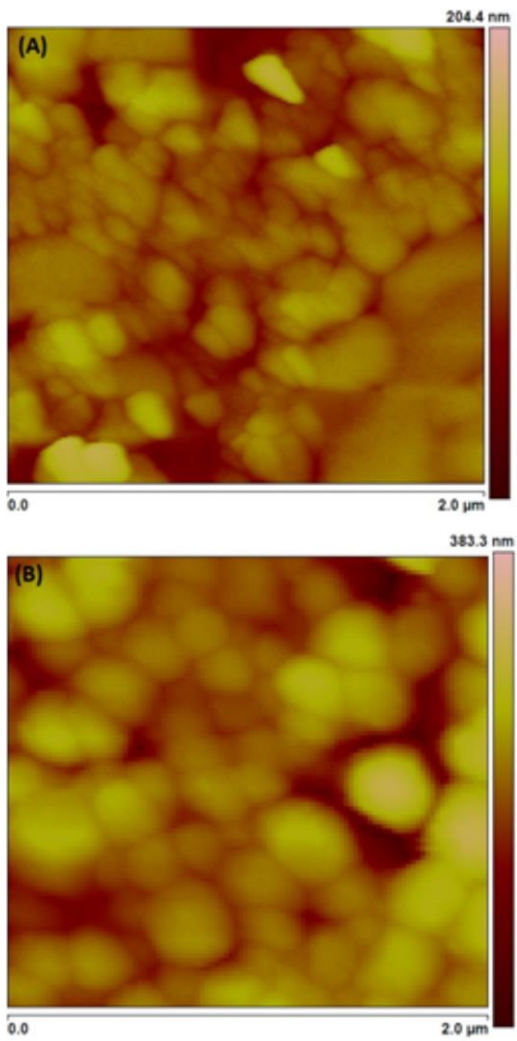

Fig. 2

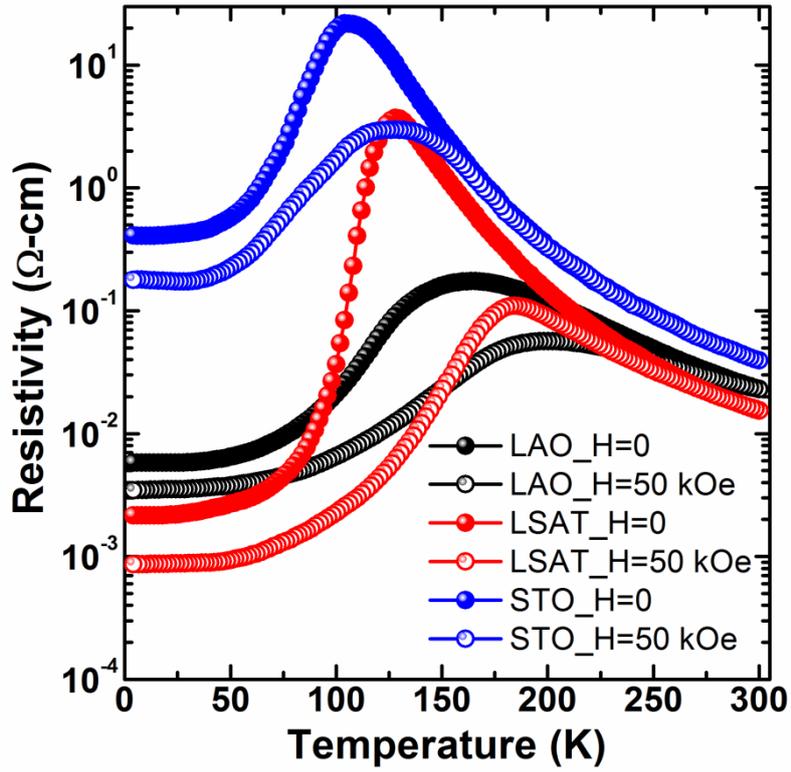

Fig. 3

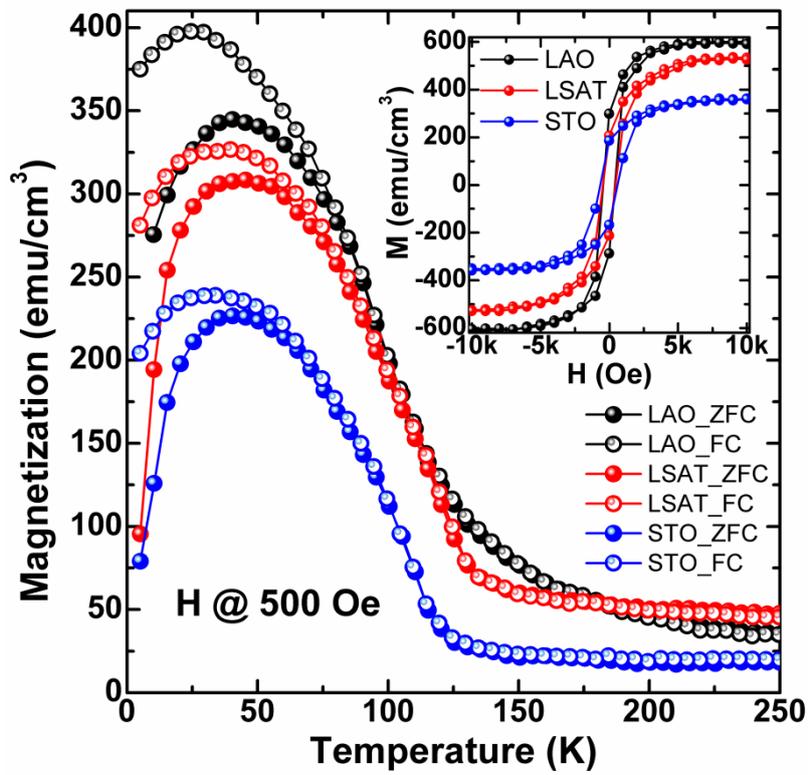

Fig. 4

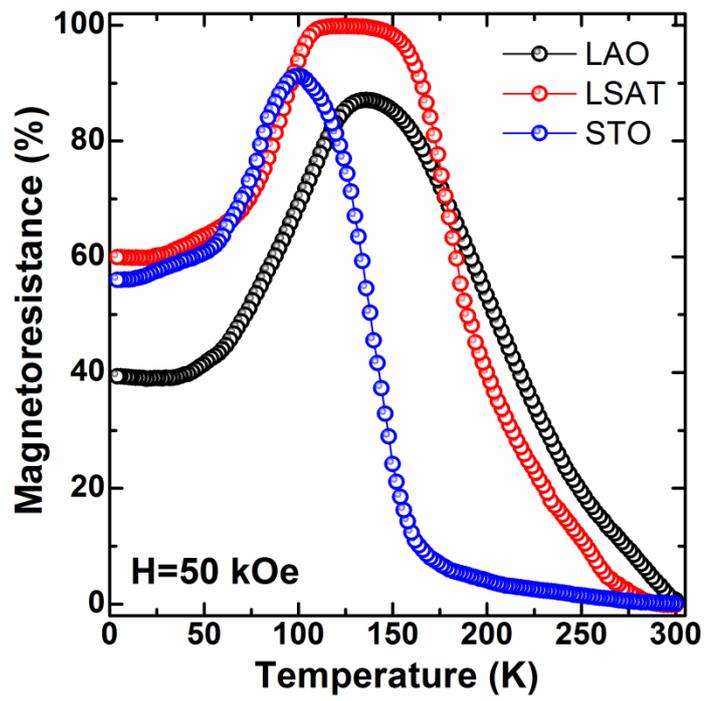

Fig. 5

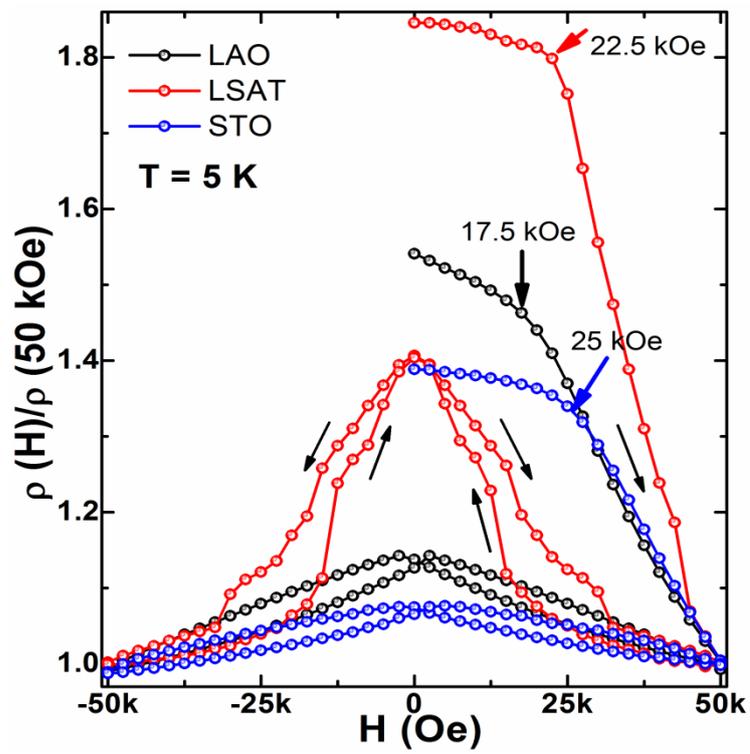

Fig. 6

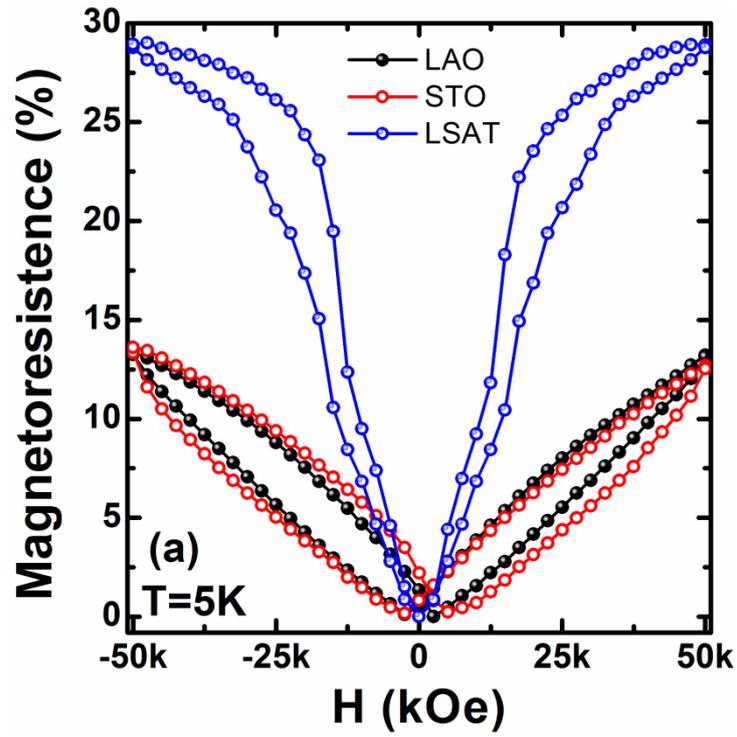

Fig. 7a

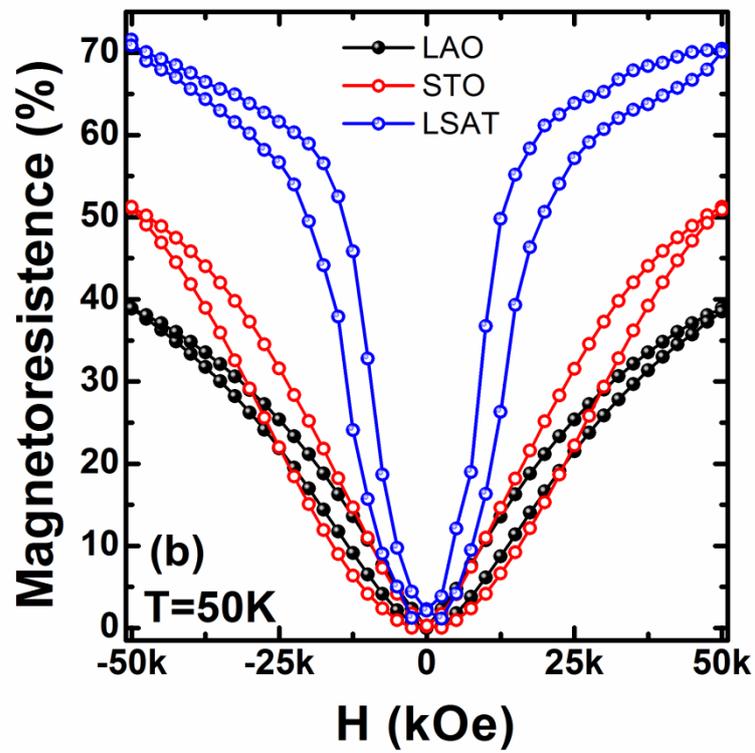

Fig. 7b

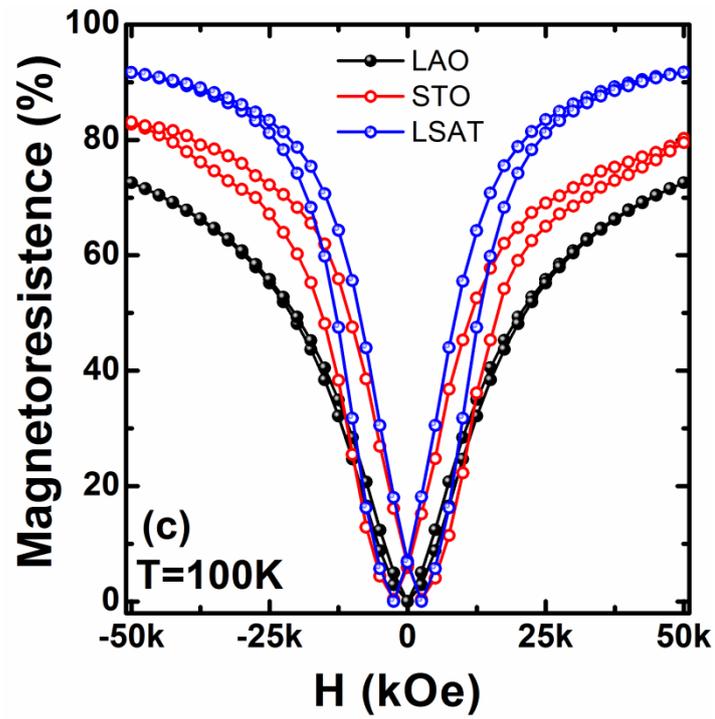

Fig. 7c

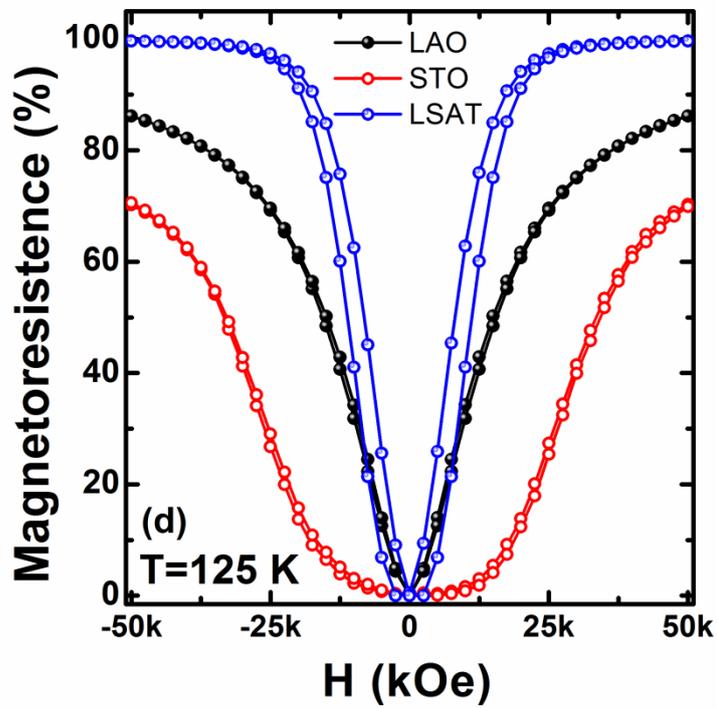

Fig. 7d

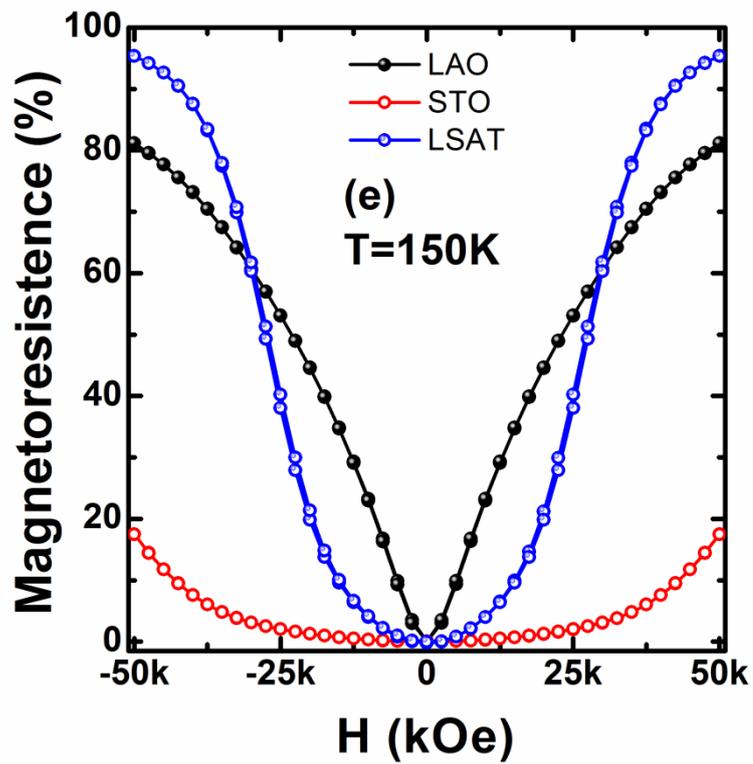

Fig. 7e

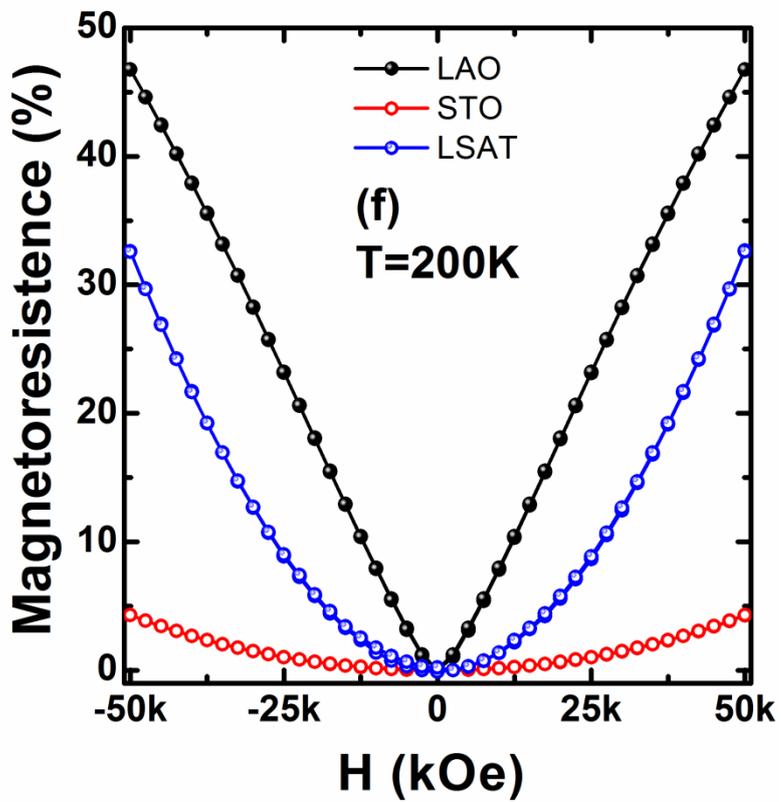

Fig. 7f